\documentclass[final,1p,times]{elsarticle}

\usepackage{amssymb}
\usepackage{amsmath}

\journal{Nuclear Physics A}

\begin{document}

\begin{frontmatter}



\title{Meson-baryon nature of $\Lambda(1405)$ in chiral dynamics}




\author[TIT]{Tetsuo Hyodo}
\author[YITP]{Daisuke Jido}
\author[RCNP]{Atsushi Hosaka}

\address[TIT]{Department of Physics, Tokyo Institute of Technology, 
	Meguro 152-8551, Japan}
\address[YITP]{Yukawa Institute for Theoretical Physics,
Kyoto University, Kyoto 606-8502, Japan}
\address[RCNP]{Research Center for Nuclear Physics (RCNP),
Ibaraki, Osaka 567-0047, Japan}

\begin{abstract}
We present the recent progress in the study on the origin of baryon 
resonances in chiral dynamics. It is shown that in the chiral 
coupled-channel approach, the form of the interaction cannot be specified 
due to the cutoff of the loop integral. To avoid this ambiguity, we propose 
a natural renormalization scheme, which affords a clue to the origin of 
the resonance, together with the phenomenological fitting to experimental 
data. We study the structure of the $\Lambda(1405)$ resonance in comparison 
with the $N(1535)$.
\end{abstract}

\begin{keyword}
Chiral dynamics\sep structure of the $\Lambda(1405)$\sep the CDD pole
\PACS 14.20.--c \sep 11.30.Rd \sep  24.85.+p

\end{keyword}

\end{frontmatter}


\section{Introduction}\label{sec:intro}

The structure of the $\Lambda(1405)$ resonance has been extensively discussed in 
strangeness nuclear/hadron physics~\cite{Weise,Gal:2008ze}. From the viewpoint 
of hadron spectroscopy, the $\Lambda(1405)$ is considered to be a meson-baryon 
molecule, since the coupled-channel scattering framework~\cite{Dalitz:1967fp} 
seems to be favored over a simple three-quark picture, while possible 
multi-quark structure is also discussed. On the other hand, the $\Lambda(1405)$ 
is of particular importance for the behavior of the kaon in 
nucleus~\cite{Akaishi:2002bg}, since this resonance strongly dictates the 
subthreshold $\bar{K}N$ interaction~\cite{Hyodo:2007jq}. Clarification of the 
structure of the $\Lambda(1405)$ resonance as well as the study of $\bar{K}$ in 
nucleus are within the scope of current/future experiments in facilities such as 
SPring-8, Jefferson Lab., and J-PARC.

It is our aim here to study the structure of the $\Lambda(1405)$ resonance from 
the viewpoint of scattering theory. We utilize the theoretical framework of 
chiral coupled-channel approach~\cite{Kaiser:1995eg,Oset:1998it,Oller:2000fj,
Lutz:2001yb}, where the excited baryons are described as resonances in the 
meson-baryon scattering amplitude, generated by the chiral low energy 
interaction. In this framework, any components other than dynamical two-body 
state (meson-baryon molecule) are expressed by the Castillejo-Dalitz-Dyson (CDD) 
pole contribution~\cite{Castillejo:1956ed}. Thus, the magnitude of the CDD pole 
contribution in the amplitude tells us about the origin of resonances. We 
present the method to extract such an information within the chiral 
coupled-channel framework~\cite{Hyodo:2008xr}.

\section{Formulation}\label{sec:formulation}

We start with the general form of the amplitude based on the N/D 
method~\cite{Oller:2000fj}, for $s$-wave single-channel meson-baryon scattering 
at total energy $\sqrt{s}$:
\begin{align}
    T(\sqrt{s};a)
    =& \left[V^{-1}(\sqrt{s})-G(\sqrt{s};a)\right]^{-1} ,
    \label{eq:TChU}
\end{align}
where $V(\sqrt{s})$ is the kernel interaction constrained by chiral low energy 
theorem. The once subtracted dispersion integral $G(\sqrt{s};a)$ can be 
identified as the meson-baryon loop function with dimensional regularization, 
and the subtraction constant $a$ plays the role of the cutoff of the loop 
function. In this way, Eq.~\eqref{eq:TChU} is regarded as the solution of the 
algebraic Bethe-Salpeter equation. The interaction kernel $V(\sqrt{s})$ is 
determined by the order by order matching with chiral perturbation 
theory~\cite{Oller:2000fj}. At the leading order, $V(\sqrt{s})$ is given by the 
$s$-wave interaction of the Weinberg-Tomozawa (WT) term 
$V(\sqrt{s})=V_{\text{WT}}(\sqrt{s})=-C(\sqrt{s}-M_T)/(2f^2)$ where $C$, $M_T$ 
and $f$ are the group theoretical factor, the baryon mass, and the meson decay 
constant, respectively. 

In the framework of N/D method, the CDD pole contribution should be included in 
the kernel interaction $V(\sqrt{s})$. The CDD pole contribution has been 
manifested in chiral dynamics by the introduction of an explicit resonance 
propagator, or the contracted resonance contribution from the higher order 
chiral Lagrangian. We however show that the loop function $G(\sqrt{s};a)$ can 
also have the CDD pole contribution~\cite{Hyodo:2008xr}. This can be understood 
from the viewpoint of the renormalization group; Once the amplitude $T$ is 
determined by experiments, then the change of the interaction $V$ can be 
absorbed by the change of the cutoff parameter $a$ in the loop function $G$. In 
this case, the CDD pole contribution in the kernel $V$ may be effectively 
transferred to the loop function $G$. Thus, in order to study the origin of 
resonances in this approach, we should make the CDD pole contribution in the 
model under control.

For this purpose, we propose the ``natural renormalization scheme,'' in which 
the CDD pole contribution is excluded from the loop function $G(\sqrt{s};a)$. 
This can be achieved by requiring (i) no state exists below the meson-baryon 
threshold, and (ii) the amplitude $T$ matches with the interaction kernel $V$ at 
certain low energy scale, based on the validity of chiral expansion. These 
conditions uniquely determine the subtraction constant in the natural 
renormalization scheme $a_{\text{natural}}$ such that 
$G(\sqrt{s};a_{\text{natural}})= 0 \quad \text{at} \quad \sqrt{s} = 
M_T$~\cite{Hyodo:2008xr}. The similar condition was proposed in different 
contexts, but our point is to regard this condition as the exclusion of the CDD 
pole in the loop function, based on the consistency with the negativeness of the 
loop function. 

This natural renormalization scheme enables us to extract the CDD pole 
contribution hidden in the loop function in the phenomenological amplitude. In 
the following, we demonstrate it in a single-channel model. First, with the 
leading order WT term $V_{\text{WT}}$ for the interaction kernel, we fit the 
experimental data by adjusting the subtraction constant $a_{\rm pheno}$ for the 
conventional phenomenological approaches. We then try to construct an equivalent 
amplitude in the natural renormalization scheme with the subtraction constant 
$a_{\rm natural}$ and the effective interaction $V_{\text{natural}}$. The 
equivalence of the amplitude is achieved by
\begin{eqnarray}
    V^{-1}_{\text{natural}}(\sqrt s) - G(\sqrt s; a_{\rm natural}) 
    =V^{-1}_{\text{WT}}(\sqrt s) - G(\sqrt s; a_{\rm pheno}) \ .
\end{eqnarray}
This equation gives us the effective interaction $V_{\rm natural}$ as
\begin{align}
    V_{\text{natural}}(\sqrt s) 
    = &-\frac{C}{2f^2}(\sqrt{s}-M_T)+\frac{C}{2f^2}
    \frac{(\sqrt{s}-M_T)^2}{\sqrt{s}-M_{\text{eff}}} ,
    \label{eq:pole} 
\end{align}
with an effective mass $M_{\text{eff}}\equiv M_T-(16\pi^2f^2)/CM_T\Delta a)$ and 
$\Delta a = a_{\text{pheno}} - a_{\text{natural}}$. The 
expression~\eqref{eq:pole} indicates that the interaction kernel 
$V_{\text{natural}}(\sqrt s)$ can have a pole, whose relevance depends on the 
scale of the effective mass $M_{\rm eff}$. This pole can be the source of a 
resonance in the amplitude, and so it is interpreted as the CDD pole 
contribution. If $\Delta a$ is small, the effective pole mass $M_{\text{eff}}$ 
becomes large and the second term of Eq.~\eqref{eq:pole} can be neglected in the 
resonance energy region $\sqrt{s} \sim M_{T}+m \ll M_{\text{eff}}$. If the 
difference $\Delta a$ is large, the effective mass $M_{\rm eff}$ gets closer to 
the threshold and the pole contribution is no longer negligible. Thus, we can 
estimate the effect of the CDD pole contribution from the values of the 
effective mass $M_{\rm eff}$.

\section{Numerical results}\label{sec:numerical}

We analyze the meson-baryon scattering and the $\Lambda(1405)$ in $S=-1$ and 
$I=0$ channel and the $N(1535)$ in $S=0$ and $I=1/2$ channel, using the method 
explained above. We use the phenomenological models~\cite{Hyodo:2002pk,
Hyodo:2003qa} to determine the phenomenological subtraction constants 
$a_{\text{pheno},i}$. These models are based on the results in 
Refs.~\cite{Oset:2001cn,Inoue:2001ip}, and the scattering observables are well 
reproduced by the interaction kernel of the WT term. The natural values of the 
subtraction constants $a_{\text{natural},i}$ are obtained by setting $G(M_N)=0$ 
for all channels.

We first evaluate the effective interaction in the natural renormalization 
scheme and extract the pole positions in the kernel. The nearest pole in each 
channel is given by $z_{\text{eff}}^{\Lambda^*} \sim 7.9 \text{ GeV}$ and 
$z_{\text{eff}}^{N^*} = 1693 \pm 37 i\text{ MeV} $.\footnote{With $n$-coupled 
channels, the effective interaction has $n$ poles, and a pair of complex poles 
can also appear~\cite{Hyodo:2008xr}.} It is observed that the pole for the 
$N(1535)$ lies in the energy region of interest, while the pole for the 
$\Lambda(1405)$ is obviously out of the scale of the physics of the resonance. 
This result indicates that the $N(1535)$ may require some CDD pole contribution, 
whereas the $\Lambda(1405)$ does not.

\begin{figure}[tbp]
    \centering
    \includegraphics[width=8cm,clip]{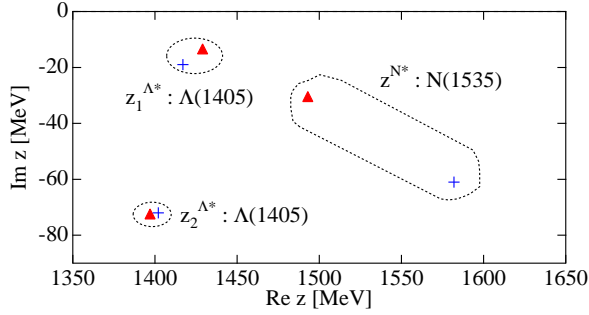}
    \caption{\label{fig:pole}
    Pole positions of the meson-baryon scattering amplitudes. The crosses 
    denote the pole positions in the natural renormalization with the WT 
    interaction (dynamical component); the triangles stand for the pole 
    positions with the phenomenological amplitude. $z_1^{\Lambda^*}$ and 
    $z_2^{\Lambda^*}$ are the poles for the $\Lambda(1405)$ in the $S=-1$ 
    scattering amplitude, and $z^{N^*}$ is the pole for the $N(1535)$ in 
    the $S=0$ amplitude.}
\end{figure}

Next we consider the idealized situation of purely dynamical components, which 
can be obtained by adopting the WT term with the natural subtraction constant. 
The pole positions in this model are shown by crosses in Fig.~\ref{fig:pole}. 
Note that in chiral dynamics the $\Lambda(1405)$ is described as the two poles 
in the complex energy plane~\cite{Jido:2003cb,Hyodo:2007jq}. These poles can be 
compared with the those in the phenomenological amplitude shown by the triangles 
in Fig.~\ref{fig:pole}, which corresponds to the physical resonances. The
phenomenologically extracted poles for the $\Lambda(1405)$ (triangles) appear 
near the dynamical ones (crosses). This indicates the dominance of the 
meson-baryon component in the $\Lambda(1405)$. On the other hand, the pole for 
the $N(1535)$ moves to the higher energy when we use the natural values. 
Although the dynamical component generates a state by itself, the physical 
$N(1535)$ requires some more contributions, which is expressed as the pole in 
the effective interaction at 1.7 GeV. Thus, the comparison in 
Fig.~\ref{fig:pole} also indicates the dynamical nature of the $\Lambda(1405)$ 
and the sizable CDD pole contribution for the $N(1535)$.

\section{Summary}\label{sec:summary}

We have studied the structure of baryon resonances in chiral dynamics. It 
is shown that the CDD pole contribution can be excluded from the loop 
function in the natural renormalization scheme. A methodology to study the 
origin of the resonance is discussed using the effective interaction kernel 
in the natural renormalization scheme, which is obtained in comparison with
the phenomenological amplitude.

We analyze the origin of the $\Lambda(1405)$ and the $N(1535)$ in 
meson-baryon scattering. In the natural renormalization scheme, the 
physical $\Lambda(1405)$ can be well reproduced with the WT interaction 
for the kernel of the scattering equation, while the $N(1535)$ requires a 
substantial contribution in addition to the WT term, especially a pole 
singularity at around 1.7 GeV. These facts indicate that the $N(1535)$ may 
not be a pure dynamical state and it requires substantial CDD pole 
contribution in its structure. On the other hand, the $\Lambda(1405)$ can 
be mainly described by a dynamical state of the meson-baryon scattering.
This conclusion is qualitatively consistent with the analysis of the $N_c$ 
scaling~\cite{Hyodo:2007np,Roca:2008kr} and the estimation of the 
electromagnetic size~\cite{Sekihara:2008qk}. 

\section*{Acknowledgement}

This work was partly supported by the Global Center of Excellence Program by 
MEXT, Japan through the Nanoscience and Quantum Physics Project of the Tokyo 
Institute of Technology, and by the Grant-in-Aid for Scientific Research 
from MEXT (No.21840026, No.20028004, No.19540297). This work was done under 
the Yukawa International Program for Quark-hadron Sciences (YIPQS).



\end{document}